# Convectivity and Rotation of Entropicly Defined Clusters As a Measure of Hurricane and Tornado State

*Bryan Kerman*[1]

*Abstract*

A statistical physical model of the two basic properties of clusters within a hurricane - their convectivity and rotation - reveals a relationship for the time evolution of a hurricane. Non-doppler data from NEXRAD surface radar imagery, of Hurricane Irma transiting over Florida, is decomposed into unique clusters on the basis of an annealing process using entropy and energy differences between pixels. Application of the concept of entropic forces between a cluster's pixels, provides an estimate of the radial velocity of each cluster by application of Stokes' theorem. The ratio of the characteristic rotation and convectivity, associated with radial flow, integrated over the extent of the hurricane, closely tracks the hurricane's state, providing more time resolution than aircraft sorties alone allow. It is concluded that monitoring the rotational and convective state, in conjunction with the size of a cluster, is capable of quickly providing forecasters and others with changes in a hurricane's state. It is also shown that entropic tornado state can be similarly described in terms of convectivity and rotation rate.

## 1.   *Introduction*

Knowledge of the extremum wind speed within a hurricane, as measured by an aircraft transit, has formed the basis for the National Hurricane Center's hurricane category. It has been successful over the years but more, is currently required (i.e. prediction of high water and storm surges and tornadoes, Klotzbach et al., 2020), which has stimulated supportive research and engineering at numerous laboratories. Another technique, originally dating from the late 1960's (Dvorak, 1984), which has been enriched over the years. It uses templates of characteristic cloud structures associated with a hurricane, to estimate changes in a hurricane's intensity.  More recently, 'deep learning', a computer-driven feature recognition process to match hurricane images with an image or numeric such as hurricane wind speed to extract such data (state, wind speed) from independent input images (e.g. Wimmers et al., 2019) has been employed. The method is quite successful in its objective, but unsatisfying in understanding the physics involved.

---

[1]   Retired; Atmospheric Environment Service of Canada; bkerman@lara.on.ca

To understand the essence of the problem, consider Fig. 1, which is a ground-based radar image of Hurricane Irma before it moved over the Florida Keys and then came ashore at the south-west tip of Florida.

There is clear evidence of distinct areas of strong convection (red) and ubiquitous rotation. The image's resolution is less than suggested by some distinct isolated pixels. This study will search for a description of both convection and rotation at pixel resolution and clusters of pixels, containing maximally about 60 pixels. Without the textural information, which in the sense of Information Theory, is connoted by local entropy variations, the human sense of localized convection and rotation would not be possible.

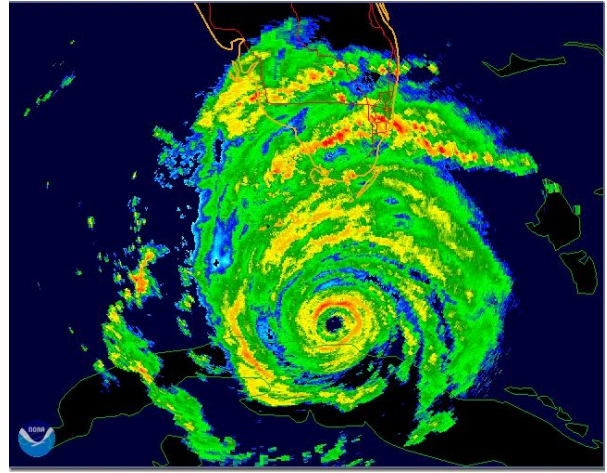

**Figure 1**: *Surface Radar Image of Hurricane Irma at 00 UDT 09/10/2017*

Another feature of Fig. 1, not necessarily obvious to the eye, is a probability distribution of 'convective intensity' (Fig. 2) that is both skewed to mostly weak convection and rare strong convection in a organized manner.

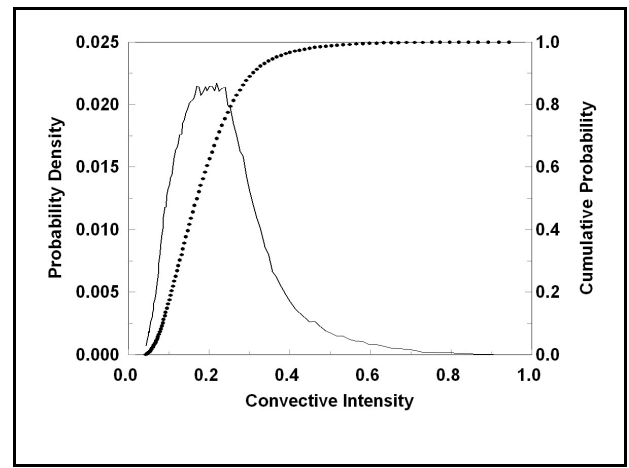

**Figure 2**: *Probability density (pdf) and cumulative probability functions of radar reflectivity in Fig. 1.*

The method below extracts hurricane state from a hurricane image, by sensing both very localized (rotational) stresses and convection from rain-related variables. The hurricane state is then derived from the ratio of rotational and convective energy within annealed clusters. A critical by-product, is the estimate of the latent heating rate, where rain is occurring, which is useful to modellers.

### *1.1    Hurricane Convectivity*

The formulation and major balances of both hurricane energy and entropy, discussed by Pauluis and Held (2002), as well as Craig (1996), Bister and Emanuel (1998), and Kleidon (2010), as well as sc, provide the foundation for a definition of hurricane convectivity.

Stating with the basics, as discussed by Pauluis and Held, consider the flow's entropy which, exclusive of external effects, is approximated by

$$S \approx L + D \qquad (1)$$

where $L$ is the latent heating rate within updrafts, and D is the dissipation rate in the turbulent rain-drop field. As a result, an imbalance between latent heating and rain rate is associated with more or less entropy in the convective field. To quantify the entropy imbalance, the parameter, µ, is defined as

$$\mu = \frac{L}{D} \qquad (2)$$

Eq. 2 has 2 limit points (0,1), when there is no latent heating, and, conversely, when the latent heating rate is at its maximum, producing as much latent heat as can be removed by the dissipation rate. Its utility is discussed below in 2 methods presented for its estimation.

### *1.2    Connectivity*

One method of describing cloud morphology is in terms of the information provided by the texture and structure of the convective fields, either by examining the connectivity of nearest neighbours, or the extended connectivity related to sinuosity. The latter ultimately involves recursive searches for paths of increasing entropy change, or boundaries/walls of convective elements. The discussion here is limited to the first, a nearest neighbour analysis, which satisfies the immediate need to characterize connectedness.

### *1.2.1 Properties of Nearest Neighbours*

A useful process (Kerman, 1998; Kerman et al., 1999) in a characterization of chaotic geophysical images by using the conditional probability of the difference of some measure of energy, $E$, between a pixel and its neighbour. Here we consider the conditional probability of neighbouring pixels with an energy difference expressed as $\delta V_T^2 / 2$. Operationally, pixels located at the 8 positions around a central pixel are examined only for negative differences, i.e. that the neighbouring energy is less

It has been found in such studies that there exists a basic statistical relationship between neighbouring intensities. The differences in remotely sensed convective intensity of radar imagery of a hurricane, converted to implied terminal velocities, $v_T$, produced the same distribution of the *informational* entropy, $\chi_{cond}\ (=-\ln p_{cond})$, which in differential form is represented (Fig. 3) as a function, $\beta(\mu)$, such that

$$\frac{\partial \chi}{\partial V_T^2} = -\beta(\mu) \qquad (3)$$

It is the creation of *physical* entropy (i.e. disorder), S, that results in a chaotic spatial structure associated with the informational entropy.

The empirical results for the slope of Eq. 3 imply that the length and time scaling differs significantly over the range of the normalized rain intensity within the imagery, which further implies that physical mixing volumes and turnover times are proportionally larger, by up to a (projected) factor of 8. The inflection point in this scaling corresponds to the critical, small convective intensities, where the scaling times and lengths associated with the cloud morphology, are changing from weak sensible heating energetics to the ultimate dominance of latent heating in strong convection.

Next consider the functional structure of $-\beta(\mu)$, not in terms of the obvious radar return strength, dBz, but in terms of a simple model for rain rate. The model concept is that a seed will grow to a droplet of size, d, and then exit the cloud. The variables $(d, g, V_T, L)$ are used to form the basis of similarity relationships for the major cloud-scale energetics. These variables also form two sets of characteristic time and length scales. The first set which describes the faster and smaller process associated with drop creation is given by ($\tau_d \approx \frac{V_T}{g}, \lambda_d \approx \frac{V_T^2}{g}$), and a second set which describes the characteristic roll-over time and spatial extent of the cloud's turbulent core is given by ($\tau_c \approx \frac{V_T^2}{L}, \lambda_c \approx \frac{V_T^3}{L}$). Another non-dimensional number, $\frac{d}{\lambda_d}$, is a measure of the efficiency of the accretion process into the surrounding turbulent field which is assumed to be 1. Random initiation occurs at a rate determined by the latent heating and the terminal velocity, $t_c$ associated with the turbulent field above the interfacial layer at the cloud base. The result is a shaft of water of length, $\lambda_D$, with an expulsion width, $d$, that is released every $\tau_c$ sec. Further the expulsion/terminal velocity, is the characteristic velocity of the turbulence in the core of the cloud. Following such a characterization, a simple model of rain rate is given by,

$$R_r = \frac{d}{\tau_C} = \gamma(\mu)\frac{gd}{V_T}(\frac{L}{gV_T}) \sim \frac{gd}{V_T}\mu \qquad (4)$$

Consider Fig. 4, which is a normalized version the rain rate on the plane of the observations, and the maximum convective intensity, and from Eq.4, $\mu$, ($\mu_{max} = 1$), where $\hat{R}_r = \frac{R_r}{R_r^{max}}$. By equating the production rate of entropy with that of the rain volume, i.e

$$\beta = \hat{R}_r \qquad (5)$$

produces a very simple method to estimate $\mu$, wherever there is a sufficiently large rain rate. It is shown below $\mu = 0.15$ that there is a distinct change in the convective dynamics and the associated rainfall.

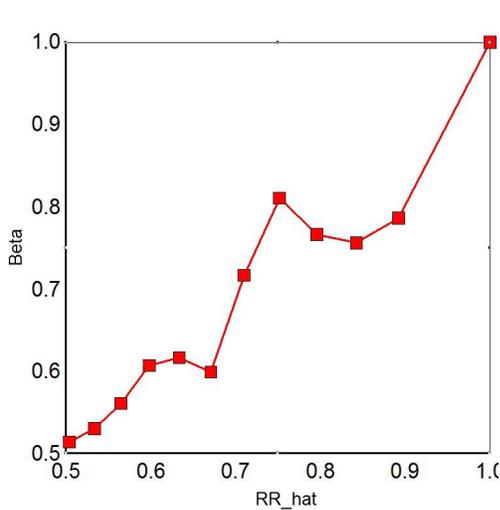

**Figure 4**: *Gibbs scaling factor, $\beta$, as a function of normalized rain rate, $R_r$.*

Accordingly by normalizing measurements of $\beta$ by its maximum observed value, $\beta_{max}$, and comparing it the normalized rain rate, associated with the maximum observed rain rate, it is clear, as seen in Fig. 4 that

$$\frac{\beta}{\beta_{max}} = \hat{\beta} = \hat{R}_r \qquad (6)$$

and hence the convectivity is also computable through its own spatially varying (Gibbs) distributions of entropy. Moreover Eq. 6, in conjunction with Eq. 5, represents another simple dynamical linkage: between cluster dynamics ($\mu$) and spatial structure within clusters.

## *2.  Hurricane Clusters*

### *2.1  Simulated Annealing*

While a radar senses areas of precipitation, it does not inherently identify organized cloud areas. A technique which examines the logical arrangement of a reflectivity field and sorts out mutually connected clusters is discussed next. Simulated annealing (e.g., Press et al.,1989 for basics) computationally mimics the physical process of metallurgical annealing in which heated metals are cooled sufficiently slowly to achieve an optimally desired `temper`, or fine grain arrangement. Sometimes called computational annealing, the only `computations` are logical comparisons of any neighbouring variable related monotonically, and closest to it, in 'temperature'. The objective for building clusters is to attain an equivalent slowest 'cooling' rate by attaching an unattached 'coolest' i.e. a minimally convective neighbour.

The computation plan identifies locations by their 'temperature' ie convective energy state, and places them in descending order in a `cluster map. If they are placed without a previously selected neighbour, they become the seed for a new cluster. If the location of the selection is adjacent to an already collected pixel, they are added to that cluster, thus allowing an existing island (cluster) to `grow`. Only when no unattached pixels surround the existing clusters, does the annealing end. It is reasonable in an analysis of a the cluster properties of convectively dominated hurricane to index them to the spatial average of either $\mu$ or $\hat{R}_r$, which relate the cluster to its energetics and its rain properties.

## 2.2 Cluster Properties

Consider a radar scan from Hurricane Irma (Fig. 1) which contains about 128k locations/pixels indicating precipitation. The annealing process results identify about 12k connected cluster pixels with sufficiently convectivity to be dominated by latent heating, i.e. about 1% of the total locations raining. In addition, the clusters are distributed as shown in Fig. 4 which indicates that many clusters are less than about 8-10 pixels in size. The first question about these clusters is 'Are they real physical identities?'. Proof that they are follows from their observed negative exponential form of their probability distribution for sufficiently large clusters (above 10 pixels), (Fig. 5) which is similar to observed cloud size distributions (Simpson, 1972). The second proof is their coherence when isolated in radar images at subsequent times, and at neighbouring heights. The third and most significant reason to believe that the organization that connected them will be related to structure arising from the convection.

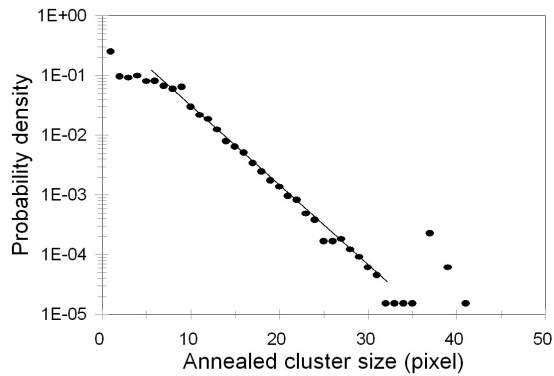

**Figure 5:** *Size distribution of annealed clusters.*

### 2.3    *Implied Cluster Forces*

Consider next the entropic force field around pixels attached within a cluster, and ultimately their sum over a cluster. Entropic forces (Wikipedia) are defined in terms of the gradient of the entropy of a field, recognizing that forces within a system will be in the direction of entropy maximization. As noted in these references, the concept of entropic forces, and more recently entropic information, has been successfully applied in studies as varied as astronomy, particle and material physics, and more recently as a possible physical manifestation of 'artificial' intelligence (Wisser-Gross and Free, 2013). It appears that there is no referenced report of the approach being applied in atmospheric science.

An entropic force is any acceleration driven by an entropy gradient weighted by some measure of 'temperature'. The entropy to be studied here, for its implied accelerations, is that (ultimately summed) within a cluster. Differences in the pixel's entropy are measured between pixel centres on a 9 x 9 grid. That entropy, $S = E/\mu$, is stored in two measured components ($\mu$ and $V_T$), where the energy and the entropy are calculated by $E = V_T^2/2 \longleftrightarrow S = E/\mu$. From the definition of an entropic force (Wikipedia)

$$\vec{a}_e = \mu \frac{\partial S}{\partial \vec{x}} \qquad (7)$$

and by expanding S, the acceleration is then written as

$$\vec{a}_e = \frac{\partial E}{\partial \vec{x}} - S\frac{\partial \mu}{\partial \vec{x}} \qquad (8)$$

and, on expansion of E and S, the relationship for entropic forces within and on the perimeter of a clusters's pixels becomes

$$\vec{a}_e = \frac{\partial V_T^2}{\partial \vec{x}} + V_T^2 \frac{\partial \ln \mu}{\partial \vec{x}} \qquad (9)$$

## 3. *Rotational Clusters*

The first term in Eq. 9 represents irrotational accelerations associated with divergence and convergence of energy, and integrates to 0 around any closed path (here the edge of a cluster) no matter the details of the path. The second term evaluates the shear as the gradient of log. convectivity, weighted by the mean square terminal velocity at the location. The net rotation of a cluster is found by summing shear accelerations over a cluster - which by Stoke's theorem, is equivalent to summing the net shear acceleration, $a_{net}$, around the cluster's perimeter. The rotation rate then follows from

$$\omega = a_{net} / rad_{eff} \qquad (10)$$

The irrotational acceleration of Eq. 9 is associated with a self-organized process driven by radial (convergent/divergent) forces and the rotational acceleration is associated with cluster rotation. The direction of the divergent force is centrifugal, directed towards the cluster's perimeter, and the non-divergent component is directed towards the cluster's core.

The average irrotational horizontal acceleration (see below) for the 3 convective intensity ranges discussed in conjunction with Fig. 2, taken over a pixel length of 125 m, are 5.4 $10^{-3}$, 1.4 $10^{-2}$, 2.2 $10^{-2}$ m/sec$^2$. As cluster size increases, the rotational accelerations remain a fraction of first order vertical accelerations within the cloud's convective core which are O(g). The statistical distribution of the absolute value of these accelerations, $\vec{a}_H$, follows a basic gamma distribution; its scale', $\hat{a}_H$, and 'shape', $\vec{k}_H$, parameters, for the full range of convective intensity, and are displayed in Fig. 6. Of special note is the wide range of convectivity, $\mu < 0.6$, associated with a region in transition between sensible and convective energy generation, where the shape parameter, $\hat{k}_H$ and the scaling is a monotonic function of $\mu$. It is concluded that similarity extends to the statistical occurrence of horizontal accelerations which fit a universal gamma distribution of the form $a_h \exp-(a_h / \hat{a}_h)$ over the bulk of clusters in this range of convective intensity.

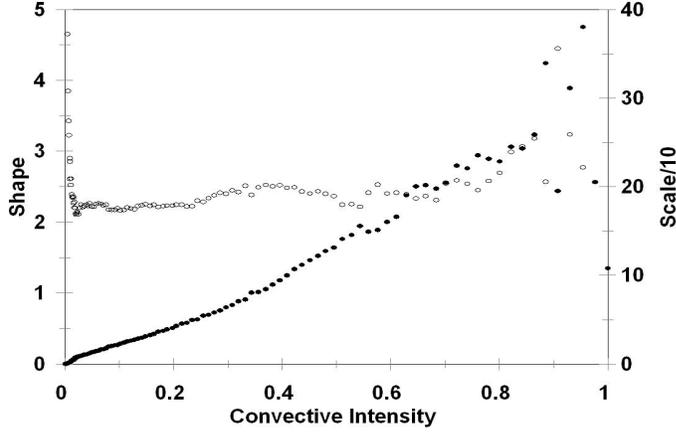

**Figure 6.** *Shape and scale parameters of a gamma distribution as a function of cluster convective intensity.*

Consider next the secondary irrotational vertical flow, in particular its Bernoulli pressure drop caused by the (first order) convective velocity in the core

$$\delta \vec{p}_z = \frac{1}{A_{xy}} \frac{\partial V_T^2}{\partial \vec{z}} \tag{11}$$

where $A_{yz}$ is the common area of a horizontal face and the derivative in the vertical direction. Consider next the result of applying a vertical divergence operator, $\partial/\partial \vec{z}$, and, then invoking incompressibility at the time scales involved with the process, such that

$$\frac{\partial}{\partial \vec{z}} \delta \vec{p}_H = -\frac{\partial}{\partial \hat{x}} \delta \vec{p}_z = -\frac{a_v}{A_{yz}} \tag{12}$$

where $A_{yz} = \Delta_y \Delta_z = (\Delta_x)^2$. From Eqs. 11 and 12, we obtain the estimate for the resultant local secondary vertical acceleration

$$a_v = -\Delta x \, \nabla_H V_T^2 \tag{13}$$

This estimate of the second order, induced vertical acceleration associated with the cluster rotation in terms of the scale-weighted, local horizontal divergence of the terminal velocity field across the cluster is both conceptually simple and directly computable from the imagery. The vertical acceleration also provides insight into the spatial arrangement of the modulation of first order convection in the core of clouds/clusters as discussed below.

Fig. 7 displays the probability distributions of vertical pixel acceleration within clusters at 3 convective intensities ranging from 0.1 to 0.7. The case designated by ($\mu = 0.1$) represents all very weakly convective pixels; ($\mu = 0.5$) represents the distribution of vertical acceleration with modest convection, and ($\mu = 0.7$) represents significant convection. As expected, the distributions have a larger average positive vertical acceleration, which broadens with increasing $\mu$, and skews towards positive convective acceleration. While such a change with increasing convectivity is to be expected, the range of negative acceleration also increases below a critical $\mu$, and is consistent with return flow and self-organized convection.

## 3.2 Internal and Peripheral Accelerations of Hurricane Clusters

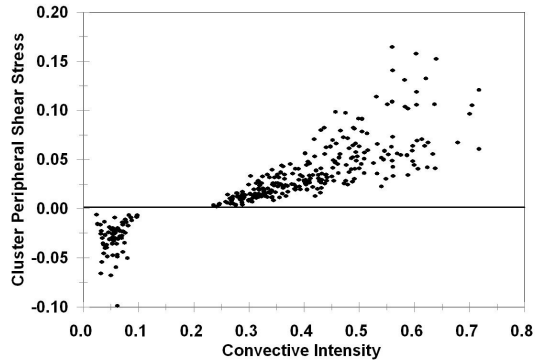

**Figure 8**: *Sensitivity of strength and direction of rotational stress on the periphery of annealed clusters ito the convective intensity.*

The net shear stress, $\sigma_{cl}$, acting on the periphery of a cluster is found by summing the local (horizontal) rotational acceleration around each pixel in a cluster using the weighted gradient perpendicular to that pixel face. The process of summation cancels opposing rotation of adjacent pixels, and only retains the sum of unbalanced estimates of horizontal flow on the periphery. The result is the magnitude of the shear stress and the implied direction of rotation on the cluster's circumference. The sensitivity of the shear stress to convective intensity is shown in Fig. 8. Also note that counterclockwise rotation exists only for, $\mu > 0.15$ i.e. significant sensible heating. Although the associated rotational speed increases with the convective intensity, so does the variability.

By isolating some of the large test clusters, $N_{pixel} > 35$, $\mu > 0.6$ and examining the locations of vertical acceleration within the cluster by eye, it is apparent that all such clusters have within them convective and subsidence sub-clusters, and that there is a correlation between convective (positive upward) acceleration and positive (counterclockwise) rotation as expected. It appears that the core of very energetic clusters is convective with subsidence localized in the outer sub-clusters. This picture of the internal cluster organization is compatible with the results of Craig and Mack (2013) of self-organized structure at second order. Another test of how well-organized a cluster is at second order, is whether the sum of the opposing secondary accelerations over a cluster area is more nearly 0 than each of the component accelerations. Fig. 9 presents a comparison for about 19 larger (>35 pixels) with significant convection, $\mu > 0.6$. There is a well defined tendency for a correlation between total subsidence acceleration and total convective acceleration in a large cluster. In addition, their sum is noticeably smaller than each component, and uncorrelated. A convenient measure of a cluster's imbalance, $h$, in terms of the positive and negative ($a_p, a_n$) vertical accelerations is

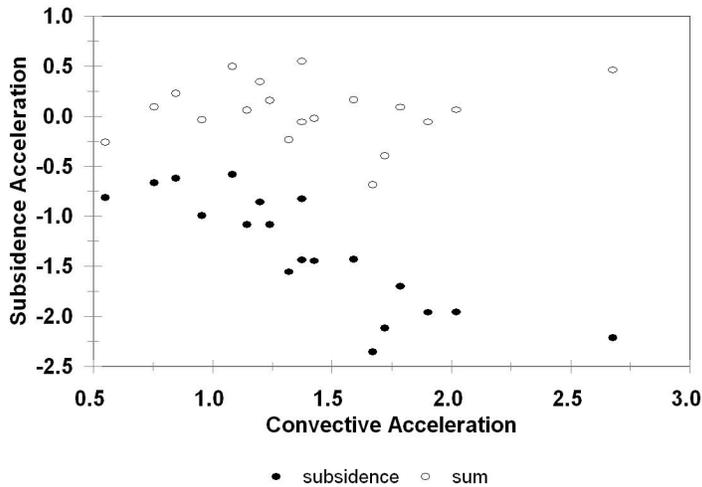

$$h = \frac{a_p + a_n}{a_p - a_n} \quad (13)$$

*Figure 9*: *Subsidence and total convective acceleration*

It is apparent that the self-organization process involving irrotational secondary vertical accelerations within precipitating columns is arranged in areas of secondary convection and subsidence to achieve optimally minimal net vertical acceleration i.e. to reduce any hydrostatic imbalance. The shear stresses of the pixels (Fig. 10) has two distinct properties – about 60% of them have negligible rotation, and, of those indicating rotation, there is a distinct preference for counter-clockwise direction (Fig. 10). The average ratio between the probability of counter-clockwise and clockwise directions for the clusters is about 1.6. The net shear stress, found by integrating local pixel rotational stresses, $\sigma_{cl}$, is equivalent to a rotational acceleration given by

$$\omega_{cl} = 2\pi \left(\frac{\sigma_{cl}}{r_{cl}}\right)^{1/2} \text{sgn}(\sigma_{cl}) \qquad (14)$$

where the equivalent radius is defined as $r_{cl} = (A_{cl}/\pi)^{1/2}$ and $A_{cl}$ is the cluster area.

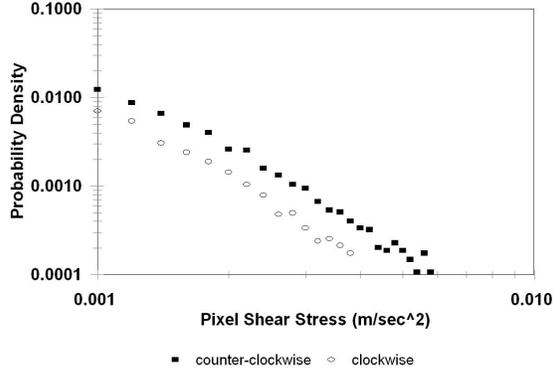

**Figure 10:** *Pdf of pixel directional shear stress.*

The pdf of cluster shear stress is markedly different (Fig. 11) than the pdf of pixel shear stress (Fig. 11) in that it is 98% positive and about an order of magnitude larger than the pixel samples. The existence of negative rotation within the pixel field, and of only positive rotation clusters, is related to the existence of organized sub-clusters as discussed next.

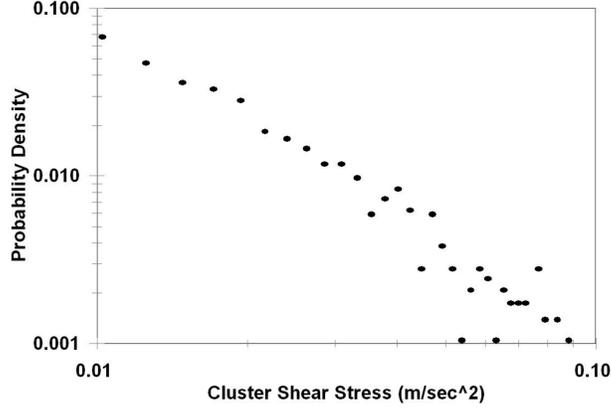

**Figure 11:** Probability of a cluster's shear stress

The net shear stress, found by integrating local pixel rotational stresses, $\sigma_{cl}$, is equivalent to rotational acceleration of a cluster given by

$$\omega_{cl} = 2\pi \left(\frac{\sigma_{cl}}{r_{cl}}\right)^{1/2} \operatorname{sgn}(\sigma_{cl}) \tag{15}$$

where the equivalent radius is defined as $r_{cl} = (A_{cl}/\pi)^{1/2}$, and $A_{cl}$ is the cluster area.

## 4. *Relationship of Energetics and Rotation in Convective Clusters*

### *4.1 Dimensional Analysis*

A simple model of the dynamical aspects of a cluster (rotation and vertical acceleration) is suggested by dimensional analysis of latent heating. If we postulate that the length and time scale for a cluster is the effective radius and its rotation rate, then by dimensional analysis

$$[L] \to \frac{E}{t} \to \frac{v^2}{t} \to \frac{l^2}{t^3} \sim (r^2 \omega^3)_{cl} \tag{16}$$

suggests the basic linkage between a cluster's latent heat production and its rate of conversion to rotational energy. Also, by the same argument, the vertical acceleration, $a_v$, by dimensional analysis, is related to $L$ by

$$[L] \to \frac{l^2}{t^3} \sim a_v \, v_r \sim (a_v r \omega)_{cl} \qquad (17)$$

which is the product of (cluster) vertical acceleration and radial velocity. Also, by equating the two Equations, 16 and 17, the vertical acceleration is expected to be of the order of the horizontal acceleration, i.e.

$$a_v \sim r_{cl} \omega^2 = \frac{(r_{cl}\omega)^2}{r_{cl}} = \frac{v_{rad}^2}{r_{cl}} \qquad (18)$$

## *4.2 Angular Momentum*

The conceptual model which arises from the existence of counter-rotating, subsiding sub-clusters in a cluster is that there is a contest for angular momentum between the counter-clockwise rotating cluster and the clockwise sub-clusters. Accordingly the observed angular momentum of a cluster is the sum of the positive and negative momentum, that is

$$(\omega r)_{cl} = (\omega r)_p + (\omega r)_n \qquad (19)$$

and is less than $\omega_p r_p$ generated by the latent heating alone. It is reasonable to assume that the source of negative rotation is the dissipation rate as it removes ultimately rotational energy. If we consider the latent heating and dissipation within both the positive and negative rotating areas, an estimate of the ratio of positive and total angular momentum, defined as $\overline{R}_{mom}$, is

$$\overline{R}_{mom} = \frac{(L_p r_p + L_n r_n)^{1/3}}{(D_p r_p + D_n r_n)^{1/3}} = (\frac{r_p}{r_n})^{1/3} (\frac{L}{D})^{1/3} = (\frac{r_p}{r_n})^{1/3} \mu^{1/3} \qquad (20)$$

where p and n denote the positively and negatively rotating areas, of effective radius, $(r_p, r_d)$, of energy generation, $(L_p, L_d)$, of energy removal, $(D_p, D_d)$, and of total energy rates within the cluster of $(L, D)$. The net positive angular momentum of the cluster is $\omega_p r_p$, and the total angular momentum available for removal is, $\omega_p r_p - \omega_n r_n$, leading to the relationship

$$\frac{\omega_p r_p}{\omega_p r_p - \omega_n r_n} \sim \frac{\omega_p r_p}{2\omega_p r_p - \omega_d r_d} \sim (\frac{r_p}{r_d}) \mu^{1/3} \qquad (21)$$

The hypothesis that the ratios of momentum are related to the convectivity was tested for 14 clusters, of size $N_{pix} > 25$, and hydrostatic imbalance, $h < 0.12$. The ratio of estimates from direct calculation involving angular rotation rates of the positive and negative sub-clusters, and the cluster's convective index are shown to be approximately equal in Fig. 12. The best fit ratio is 0.66 +/- 0.026, which is both of the order of 1, and has small variability over the range of $\mu$.

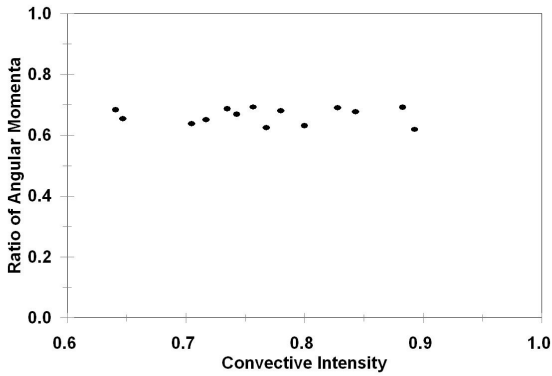

**Figure 12:** *Ratio of positive and negative momentum as function of convective intensity.*

An empirical fit of the fraction positive angular momentum associated with latent heating to that associated with dissipation, as $\mu \to 1$, is given by,

$$\frac{\omega_{cl} r_d}{\omega_p r_p} \approx 0.5 + 0.67(\mu - 0.6) \to 0.9 \qquad (2$$

2)

which provides an empirical estimate for the relative size of the positively and negatively rotating sub-clusters, given by

$$\frac{r_n}{r_p} \approx 0.5 + 1.25(\mu - 0.6) \qquad (23)$$

for clusters with negligible hydrostatic imbalance, $|h| < 0.055$. Together, they provide a model for the size and rotation rates of the positive and negative sub-clusters as a function of $\mu$. Interestingly, Eq. 23 predicts that the negatively rotating sub-clusters increase to match those positively rotating in the limit where the production matches the dissipation of energy, that is $r_n \to r_p \to 0.5$ as $\mu \to 1$. The explicit effect of a larger hydrostatic imbalance on these relationships could not be found, except to note that it appears to be significant.

## 5. *Height Dependence of Energetics and Cluster Rotation*

The previous analysis of the properties of clusters in a hurricane was limited to a single height. Much is known of the structure, both horizontally and vertically, within such structures. What is not understood is the interplay between individual clouds in determining the spatial average of the field of convection. The results of the descriptions in previous sections, of some of the energetics and rotation properties of clusters, lead to the basic question: what is the net effect, both horizontally and vertically, of the localized semi-coherent, convective process, occurring randomly in its initiation time, and location?

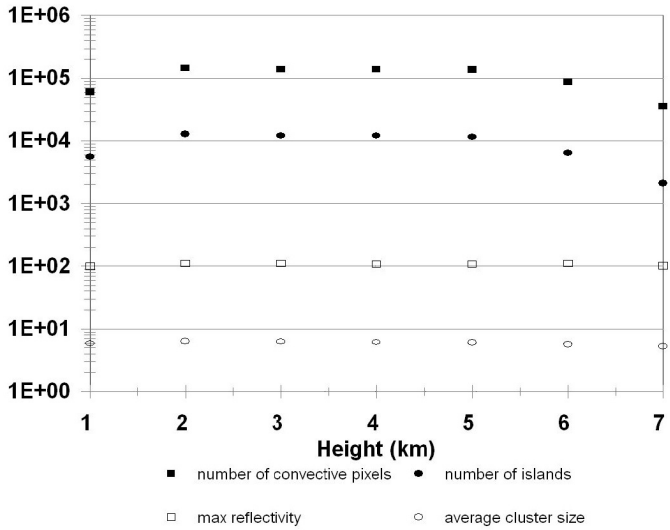

**Figure 14a**: Number of convective pixels, islands, reflectivity and average cluster size with height

The procedure, which was followed, first annealed the Irma hurricane image, described above, at 7 levels from 1000 to 7000 m. Various statistics of the overall field at both the pixel and cluster scale were assembled and are described next. A location within an image is described as convective if its convective intensity, $\mu_{cl} \geq .5$, the threshold for latent heating to exceed sensible heating, as described in Section 2.2. Some basic results are provided in Fig. 14a for the vertical structure of horizontal averages at a given height of the number of sufficiently convective pixels, of the number of annealed clusters, of the maximum reflectivity and the average size of all clusters.

The most striking result in Fig. 14a,b is the presence of the nearly constant structure for the relevant process parameters - maximum reflectivity and cluster average size, particularly from 2 to 5000 m. Their near constancy implies that the convective process, despite its localization in clusters, is vertically coherently, and that image properties observed at 2000m are almost identical to the same property up to near maximum cloud-top levels. This result also offers confirmation of the consistency of the annealing process to identify clusters in that it returns almost the same number of clusters and the same average cluster size, at each level, through this core of the hurricane's convection.

Fig 14b provides other detailed properties of the cluster structure, specifically the average latent heating and dissipation rates, as well as vertical acceleration and peripheral stress. Again they are all essentially invariant through the convective core, with small decreases at the lowest and highest levels.

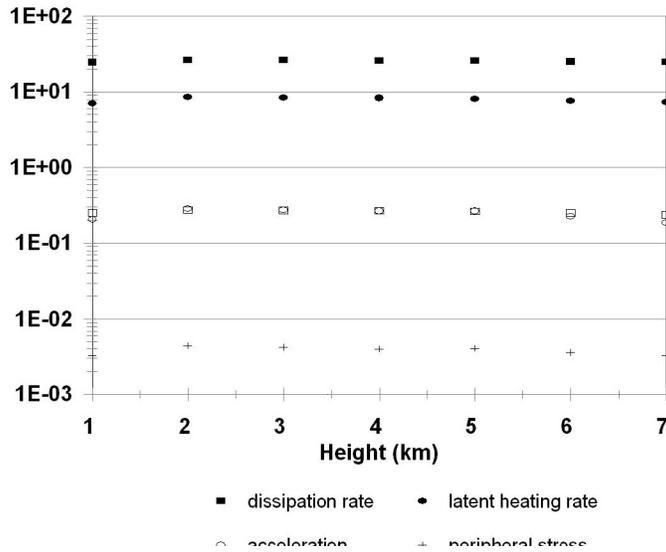

**Figure 14b**: Dissipation, latent heating, vertical acceleration and peripheral stress with height

It could be argued that the constancy is related to the constancy of the reflectivity because basic similarity can be traced to it. However that could not explain the spatial arrangements of energy and convective intensity which define the entropic force field. For the average vertical acceleration and shear stress/rotation rate to be the same vertically requires that the derivative field around pixels is solely a function of the reflectivity at that pixel which is exceedingly implausible. It is concluded au contraire that the constancy of both the reflectivity, energetics and rotation are constant through the convective core as part of a basic property of the cluster field.

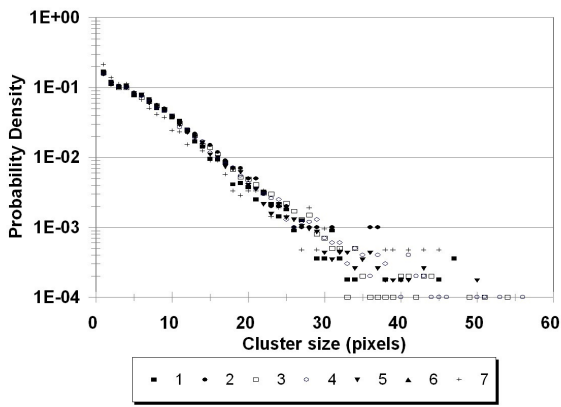

**Figure 15:** Probability of cluster size at a given height.

Clearly the structure within clusters at different heights is related to the constant convectivity layer. Accordingly, the probability density distribution of the clusters at each level was computed and displayed as a function of height (1-7 km) in Fig. 15. A logarithmic axis has been chosen based on the argument presented in Sect. 4.

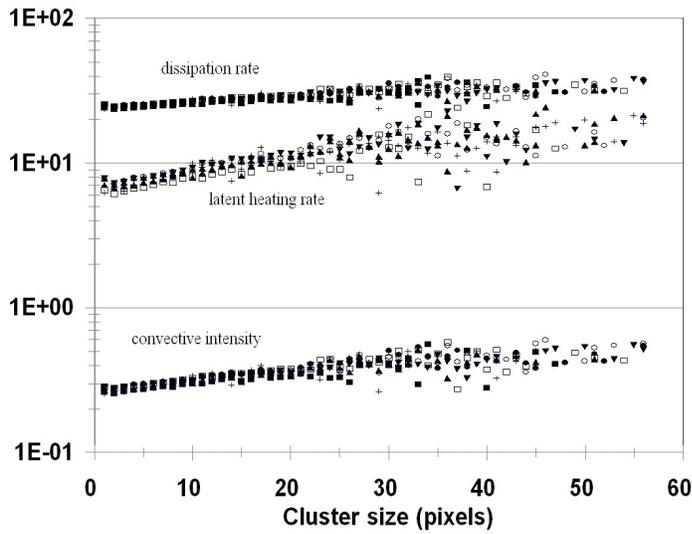

**Figure 16a:** Variation of dissipation, and latent heating rates and convective intensity as a function of cluster size.

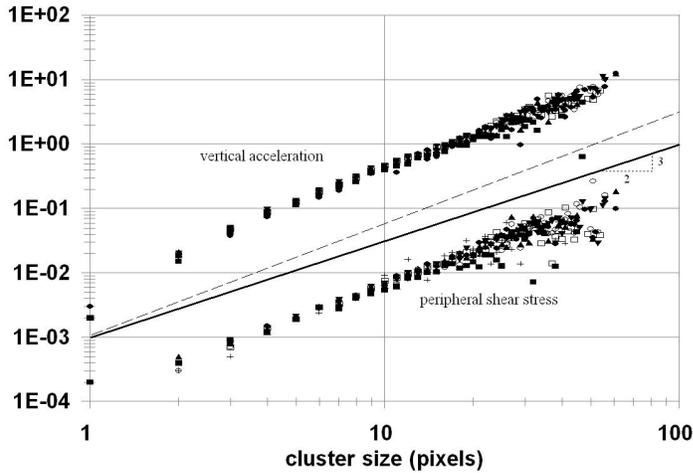

**Figure 16b:** *Variation in the vertical acceleration and peripheral shear stress with cluster size.*

All three parameters decrease with larger clusters as might be expected. The variation of dissipation, based solely on estimates of terminal velocity, increase with a constant slope. However the slope of the convectivity changes noticeably for clusters greater than about 30 pixels (0.7 km$^2$). The same break point occurs in the distribution of latent heating indicating that the change is associated with errors in the estimation of $\mu$ and/or its use in the annealing process. Fig. 16a also demonstrates that the convectivity is increasing as a result of increasing latent heating, and not decreasing dissipation, and that the break in $\mu(A)$ stems from the reduced average latent heating in large clusters. As mentioned above, this effect may not be real, but rather some feedback error associated with the clusters, annealed under local pixel convectivity, $\mu$, which may be slightly too large.

Fig. 16b compares the vertical acceleration and peripheral shear stress with cluster sizes. In this case, there is a different sensitivity, with cluster area based on linearity of the log-log plot, to a power law relationship, i.e.

$$a_v \sim \sigma_{ci} \sim A_{cl}^\gamma \qquad (23)$$

where $\gamma \approx 1.7$, based on the entire range of cluster areas. Accordingly the key dynamic factors, vertical acceleration and shear stress, increase faster than the rate of increase of cluster area. Apparently large convective clusters have an internal structure such that they can alter their vertical acceleration and rotation speed. An argument for a $A^{3/2}$ structure of Eq. 24 is readily derived from a dimensional analysis given by

$$(a_v, \sigma) \to lt^{-2} \sim A^{1/2}\omega^2 \sim A^{3/2} \qquad (24)$$

if $\omega \sim \eta A^{1/2}$ where $\eta$ is scale independent. To be definitive about the actual sensitivity to cluster size requires evaluating the annealing algorithm for systematic overestimation of $A$ as mentioned above. Whatever the exact rotation rate sensitivity to scale, the fact that it exceeds linear, argues for an unstable cycle of yet more vertical growth, more entrainment, a larger cross-sectional area, and a further increase in rotation rate with cluster area growth

## 6.    *Entropic Hurricane State*

Hurricane category, usually estimated on the basis of wind speed and eye pressure, reflects the synoptic energetic state of a hurricane. (Similarly, tornado category is based on estimated sustained wind speed of the vortex.) Next a method is sought which indicates the current state of an evolving hurricane or tornado (Appendix 4) on the basis of the mesoscale energetics of its internal rotating structure.

By extension, two non-dimensional balances involving rotating cluster energetics are required to represent a hurricane (or tornado's) internal cluster state. In addition to measurements of wind speed and pressure, the notion of the texture of the cloud field is also cited in recognizing a hurricane's category. Here the concept of texture is shown to be suitably represented by the size distribution of the clusters constituting a hurricane/tornado (Appendix 4).

In accordance with the previous developments, the convective state of a cluster is best represented by its convective intensity, $\mu$, which characterizes the relative rates of energy exchange involving latent heating, $L$, and dissipation, $D$, (in the terminal velocity field), i.e.

$$\mu = L / D \tag{25}$$

It is further suggested that a natural parameter to characterize rotation of a cluster is the 'Rossby number', the ratio of the momentum involved in the angular velocity of a rotating cluster, and the terminal velocity of the field of the precipitation within the cluster, i.e.

$$\psi = \frac{r_{eff}\,\omega}{V_T} \tag{26}$$

In addition, a simple variable which characterizes texture of an image is the scale size of a (negative exponential) distribution of cluster sizes, $A$. That distribution can be shown to be a function of the state/age of a cluster field, but is not utilized here.

These parameters are not independent. For example, as dissipation is defined by $D = gV_t$, and $(\mu, \psi)$ are both functions of $D$. In addition, both $\mu$ and $\psi$ are functions of $A$.

### 6.1. Definition of Entropic Hurricane State

As shown in Fig. 17, the ratio of the rotational speed to the average vertical terminal velocity of the precipitation within an annealed cluster, (its 'spin', $\psi$), is proportional to the cluster's ratio of latent heating to dissipation, ('convectivity'). Between 02Z and 08Z, as Irma moved away from the coast of Cuba, its intensity increased (which is also associated with an increase in large cluster convectivity). It is useful to define the net effect of spin and convectivity by forming their ratio, $\xi$, given by

$$\xi = \frac{\psi}{\mu} = \frac{r_{eff}}{V_t} \omega \frac{D}{L} = \frac{r_{eff} \, \omega \, g}{L} \qquad (27)$$

where the equivalence for dissipation, $D = gV_t$ has been substituted. $\xi$ is defined as the 'entropic hurricane state' - the ratio of spin and convectivity.

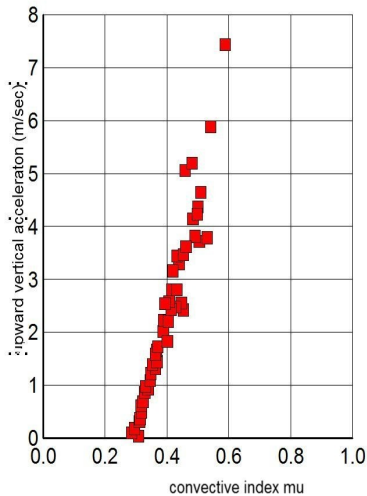

**Figure 17 Downward vertical acceleration as function of $\mu$**

## 6.2 Time Variation of Entropic Hurricane State

The variation with time of the EHS for the passage of Irma from 2017/09/10:00Z to 2017.09/11:23Z, is presented in Fig. 18. The output represents hourly averages of radar scans (at 3 km height) at about 7 minute intervals. Several labels have been placed on Fig. 18 to indicate Irma's evolving status – its hurricane category as determined by the National Hurricane Center (NHC), and the occurrence of key events.

Several features stand out. The first is that the Key West radar operating at the same time as the Miami radar indicates a higher entropic hurricane state at all common times. A plausible reason for the discrepancy is that Miami was not detecting the hurricane beyond its active core but Key West was.

Fig. 18 helps visualize the time variation of Irma, as seen by the GOES 16 satellite, at four distinct stages of its evolution. The first, 18a, occurred when Irma had moved out from the coast of Cuba, and had been measured by aircraft as a Category 4 hurricane on the NHC scale. Note its tight, closed structures with well-defined sub-features. The second image, 18b, occurred as Irma was about to move onshore between Fort Myers and Marco Island. It is somewhat larger and more diffuse than 18a. but retains its dark core. After moving very slowly on-shore and beginning its journey up the Florida peninsula (18c), Irma was considerably more dispersed, but still retained a smaller dark intense core. By the time that Irma had reached Georgia (18d), it had lost is core and the previous strong rotation, and the banding structure, had weakened considerably. The estimated hurricane state (discussed next), based not on aircraft measurements, but on ground-based radars imagery, has been placed in the figure captions of each image.

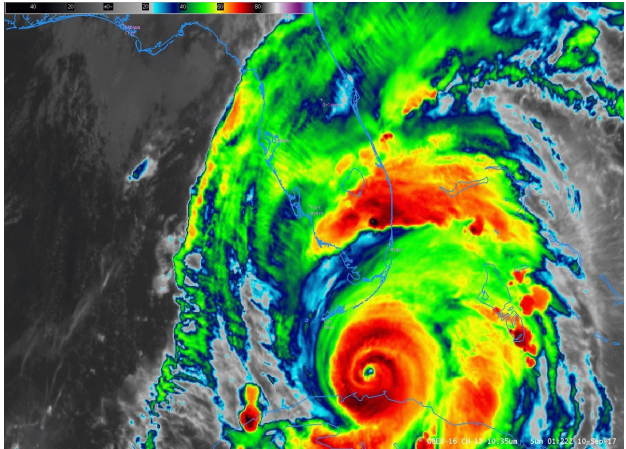

**Figure 18a:** *Irma on north shore of Cuba 2017/9/10/10Z (Cat 4)*

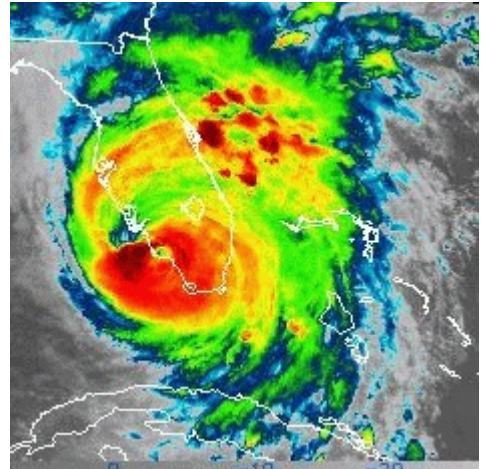

Figure 18b: Irma on shore of Florida 2017/9/10/20Z

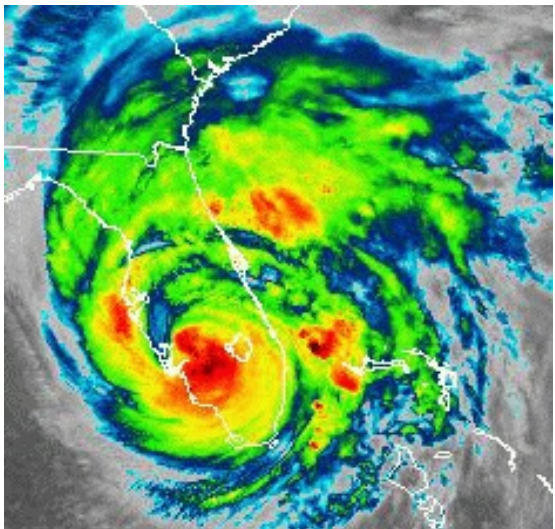

**Figure 18c**: Mid-peninsula, Florida 2017/9/11/00Z

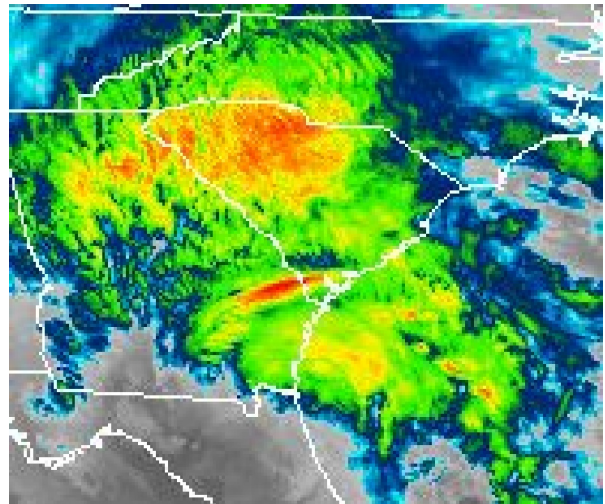

**Figure 18d:** *Dissipating, South Carolina 2017/9/11/22Z* (Tr. Stm.)

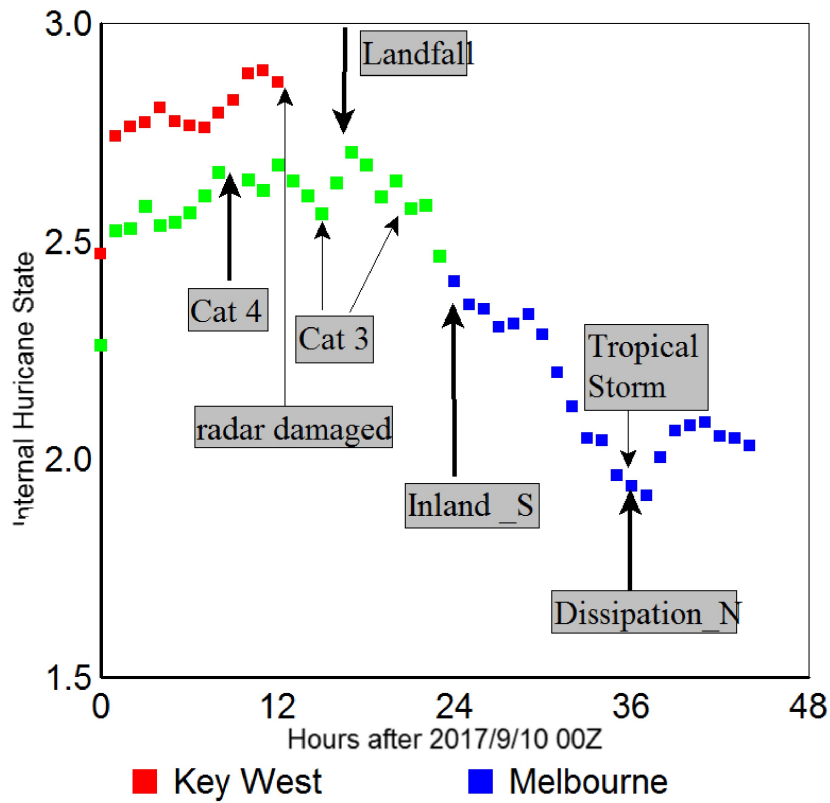

**Figure 19:** *Comparison of aircraft measured estimates of (NHS) hurricane state (Cat 3,4), with computed EHS (green, blue) (arbitrarily scaled units) min. at mid-peninsula; dissipation along axis of peninsula. Minimum at declared tropical storm. Slight recovery afterward.*

### *6.3    Observed Entropic Hurricane State*

The variation with time of the IHS for the passage of Irma from 2017/09/10:00Z to 2017.09/11:22Z, captured in Fig. 18, is presented in Fig. 19. The output represents hourly averages of radar scans (at 3 km height) at about 7 minute intervals. Several labels have been placed on Fig. 19 to indicate Irma's evolving status – its hurricane category as determined by the NHC, and the occurrence of key events.

Several features stand out. The first is that the Key West radar operating at the same time as the Miami radar indicates a higher EHS at all common times. A plausible reason for the discrepancy is that Miami was not detecting the hurricane beyond its active core, but Key West was.

It is observed that, according the EHS analysis of the Miami radar, the NHC category was actually above Category 4 when it came ashore – which was not observed by analysts as there was not an aircraft sounding at the time. Elsewhere in Fig. 19, the internal hurricane scale varies consistently with the known/physically observed NHC hurricane categories in that it produces the same estimates when the two Category 3 occurrences were observed. Lastly, the change in the NHC is dramatic (and linear) in the dissipation stage between the start, ('S') to end ('N') designations) but more subtle (but observable), with gradual changes in the NHC category estimates until it reached tropical storm status.

Note that the 2 radars – Miami and Melbourne – show no sign of a displacement in hurricane state during their common time. The difference in this case, and the first involving Key West, may be that Irma was essentially between the Miami and Melbourne radars, and presumably was sensed similarly by both radars.

From Fig. 19, it is possible to tentatively conclude that the entropic hurricane state,

a. follows the current hurricane categories observed by aircraft flights faithfully (on an hourly) basis, but

b. suffers from underestimation if the radar is not sampling the entire convective area.

As a result, it is recommended that a similar method be developed using satellite data, particularly from GOES etc infrared and AMSR micro-wave imagers, which capture the entire hurricane extent and can follow the evolving hurricane in far off-shore locations. However it remains to be proven that the computation of the entropic hurricane state using such satellite imagery is possible.

## 7. Conclusions

### 7.1 Major Results

There exists a similarity structure of the large scale energetics of a hurricane. Existing methods of deriving physical variables, particularly terminal velocity of rain, drop radius, and the rain rate, lend themselves to an evaluation of the energetics at the pixel scale. A key variable, applicable at various scales, is the convective intensity, $\mu$, which acts as 'temperature' parameter, for a similarity theory of moist convection. The convective intensity is derivable from measurements of the rain rate, or from an informational entropy process involving nearest neighbour energy differences. Simulated annealing of the hurricane radar image using the convective intensity produces a field of spatially independent rotating clusters. The energetics of a cluster is explicitly tied to its angular momentum.The application of entropic forces, first at the pixel resolution, and then integrated over a cluster, produces estimates of vertical accelerations and rotation of a cluster.
.

### 7.2 Future Research

During the course of the research reported here, several other analyses of the energy were applied and, although not encorporated, proved fruitful, as listed below:

1. Evidence exists of a modified form of Richardson dispersion relationship within the hurricane's cloud field.

2. To be able to trace individual clusters with time and height, a probit analysis involving invariants associated with the energetics and angular momentum, proved promising.

3. The influence of the overlying thermal stratification on the rotation of clusters needs to be clarified.

4. The method also successfully identified individual tornadoes in radar images and measured their energetic state, as per a hurricane, using their rotation rate and convectivity. (Appendix 4)

## *Appendix 1:*

### *Entropic Forces*

Entropic forces are the subject of evaluation in several important research topics: astronomical gravitation, atomic physics, and colloidal mixtures. However, to the author's knowledge, there has not been an application in meteorological research. As demonstrated, entropic forces prove to be both natural, and readily computable from remotely sensed imagery

An entropic force is defined as any acceleration driven by an entropy gradient weighted by some measure of 'temperature' in the sense of sub-scale, statistically defined, non-observable, turbulent energy. The entropy studied here, for its implied accelerations, is that estimated within individual pixels, and the spatial structure of these pixels within a cluster.

Consider the expression of a simplified description of entropic force in a convective atmosphere, given by

$$a_e = \mu \frac{\partial S}{\partial x} \quad (A1)$$

where μ, the convectivity, is an entropic measure of 'temperature'. It is dimensionless and not directly measurable. S, the (pixel) entropy, related to the pixel's vertical kinetic energy, $E$, is in terms of the estimated terminal velocity of rain droplets, $V_T$, (where $E = V_T^2/2$), give by the relationships

$$S = E/\mu = \mu V_T^2 / 2 \quad (A2)$$

$$L \sim \frac{\partial}{\partial t} V_t^2 \sim \frac{\partial}{\partial t} KE \quad (A3)$$

The condition of no acceleration ($S = 0$) occurs when the flow is either (or both) not convective ($\mu = 0$) and not precipitating ($v_t = 0$). Further (A3) is the condition that the condition arising for time change of the acceleration of a given sign at a pixel which, at a location of no force, reduces to

$$\frac{\partial V_T^2}{\partial x} = 0 = \frac{\partial \mu}{\partial x} \quad (A4)$$

Accordingly, there are 2 components of an entropic acceleration at a point - a classical horizontal (Newtonian) acceleration in terms of terminal velocities only, and a second (entropic) acceleration which has a direction established by the kinetic energy weighted (directional) gradient of the convective field.

## *Appendix 2*

### *Dimensional Analysis and Computation of Convectivity*

Consider a simple model of rain rate in terms of drop radius, $r$, produced at a time rate, $V_T/g$. Consider a dimensional analysis for $\mu : \mu = M(\dot{R}, r, V_t, g)$ where is the rain rate, $V_T$ is the terminal velocity, where $g$ is the vertical acceleration, and $r$ is the mean rain drop radius. A dimensionless relationship can be found among these variables, surprisingly to first order:

$$\chi \sim \frac{\dot{R} V_T}{r g} \qquad (A5)$$

In order to estimate the similarity coefficient, $\chi$, in Eq. A5, consider an argument over related time scales. If we consider the latent heating process as the rate of production of internal energy available as kinetic energy of size, $r$, at a time scale associated with raining, the process is described by the O(1) relationship above

$$L = \frac{\dot{R}}{r} \frac{v_T^2}{2} \qquad (A6)$$

Next consider the dissipation process as droplets fall, working against the convective updrafts at terminal velocity, $V_T$ given by.

$$D = g\, v_T \qquad (A7)$$

so that the convectivity (within measurement error) is given by

$$\mu = \frac{L}{D} = \frac{\dot{R}}{d}\frac{v_T}{g} \quad (A8)$$

i.e. the similarity coefficient of Eq. A5 is 0.5.

## *Appendix 3:*

### *Empirical Estimation of Convectivity and Latent Heating from Radar Backscatter*

Latent heating in clouds is important to the dynamics of atmosphere (Pauluis and Held, 2002). However estimation of latent heating of clouds by remote sensing has proven difficult. The most widely used method is based on the development of model variables associated with latent heating from radar reflectivity, and then make model estimates of sensible temperature change with time, and equate them (Chen et al., 2021 (Irma); Guimond and Reisner, (2021); House and Yuger, 2023)).

Instead a surrogate method, which employs a computational model based on other remotely sensed physical measurables is developed here. The approach utilizes the remotely sensed rain rate, reported in terms of implied rain depth with time, and build on it a simple physical model ito other variables sensed by radar - drop size and terminal velocity - to estimate latent heating. Those relationships between these variables and radar reflectivity, in terms of the reflectivity, $Z\ (mm^6\ m^{-3})$, ($dbZ = 10\ \log_{10} Z$) are:

**Rain rate,** $\dot{R}$, $(mm\ hr^{-1})$ (Observed) $\dot{R} = 4.629\ Z^{2/3}$ (A9)

**Drop size, D, (*mm*)**

The drop size depends sensitively on the nature of the cloud type (e.g. stratiform, weakly or strongly convective etc). (e.g. Bringi et al., 2003) Of interest here is hurricane (Irma) convection which, because of the extreme turbulence and the abundance of sea salt condensation nuclei, has a very limited range of droplet sizes, around $D = 1.8\,mm$.

**Terminal velocity**, $V_T$, $(m\,\sec^{-1})$

is computed from the relationship,

$$V_T = 3.78\,D^{.67} \tag{A10}$$

which, in a hurricane, where $D$, is essentially constant,

$$V_T = 5.6\,(m\sec^{-1}) \tag{A11}$$

***Rain Rate Model*** ($\dot{R}$)

Consider again the dimensional analysis of rain rate ito of these variables and latent heating, $L$, given by

$$\frac{\dot{R}}{r} \approx \frac{L}{V_t^2} \tag{A12}$$

with no non-integer exponents. By rearranging to isolate L, and noting that $\dot{R}\,r^{-1}$ has the units of a time derivative $<\partial/\partial t>$ that Eq. A12 can be rewritten symbolically as

$$\frac{\partial V_t^2}{\partial t} \sim L \tag{A13}$$

which states the rate of change of the rain field's kinetic energy arises from the release of latent heat energy upon condensation.

Consider a re-arrangement of A5 and A8 ito the dissipation rate, $D = gV_T^2$, and the 'convectivity', defined as the ratio of latent heating to dissipation, $\mu = L/D$,

$$\left\{\frac{R}{2r}\right\}\left\{\frac{V_T}{g}\right\} \sim \mu \qquad \text{A5, A8}$$

For a hurricane case, the ratio, $V_T / rg$, is essentially fixed, so that the variability of the convectivity is mostly dependent on the (scaled) rain rate. However, the similarity coefficient in Eq. 3a is not defined. It can be shown, for a large reflectivity (~ 60 dB) the estimation of $\mu$ (Sec. 2) is very near equal 1 and therefore the similarity constant is about 0.5.

## *Appendix 4:*

### *Tornado State*

Operationally, a tornado's categorization is based on the sustained wind speed of a vortex as measured by a doppler radar. Here a new description of tornado state, based on both local back-scatter and inferred rotation, detected inside a tornado, by a non-Doppler radar, is discussed. Texture is first computed using the convectivity, $\mu$, measured over a size distribution of clusters. It is suggested that each cluster can be characterized by the ratio of its rotational velocity to its inertial/terminal speed as a (mesoscale) Rossby number, given by.

$$\Psi = \frac{r\omega}{v_t} \qquad (A14)$$

and is a function of time.

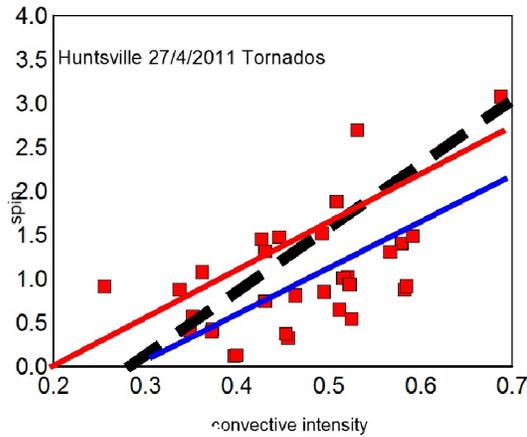

**Figure A4_1:** *Spin of tornadoes and hurricane clusters in terms of their convective intensity, $\mu$.*

An experiment was constructed to find the relationship between $\mu$ and $\Psi$ for clusters varying in size, S. Data for the study were derived from

1. the energetics of Hurricane Irma from Miami, Tampa, Tallahassee, and Birmingham along it path over several days. The time chosen for each location corresponds to the scan which had the largest number of clusters at 3 km height. Accordingly the scans characterising Irma's state as it passed each radar site.

2. the energy of a tornado rich radar image from Huntsville, Alabama on April 4, 2011 during 'Super Outbreak'.

The downloaded Nexrad images were processed identically for each data set. The latent heat and dissipation, as well as its terminal velocity. were averaged over each cluster identified during a radar scan. All variables were assigned to like-sized cluster categories, and averaged so that $\dot{L}, D, v_t, \omega, r_{eff}$ were computed as a function of characteristic size, S. Then, $\mu$ and $\Psi$ were computed for Eqs.8 and 14, and the results displayed in Figs. A4_1 and A4_2.

Several features of the computations for both hurricane and tornado state are remarkable. The first is that $\mu$ and $\Psi$ vary linearly even though there is a significant larger range of the $\Psi$ component compared to $\mu$. For reference, the black line between hurricane and tornado state, shows the variation of $\mu$ and $\Psi$ within an hourly set of scans occurring in Hurricane Irma when the hurricane category was at least 3. The other feature of the tornado data is the existence of two distinct groupings of the data, above hurricane and below hurricane strength, outlined by the red and blue lines.

The upper grouping represents clusters which have a larger spin than many hurricane clusters of the same convective intensity. The lower group represents weaker spin than observed in hurricane clusters of the same convective intensity. It is reasonable to associate the larger spin group with tornadoes and the weaker spin with either growing or decaying tornadoes.

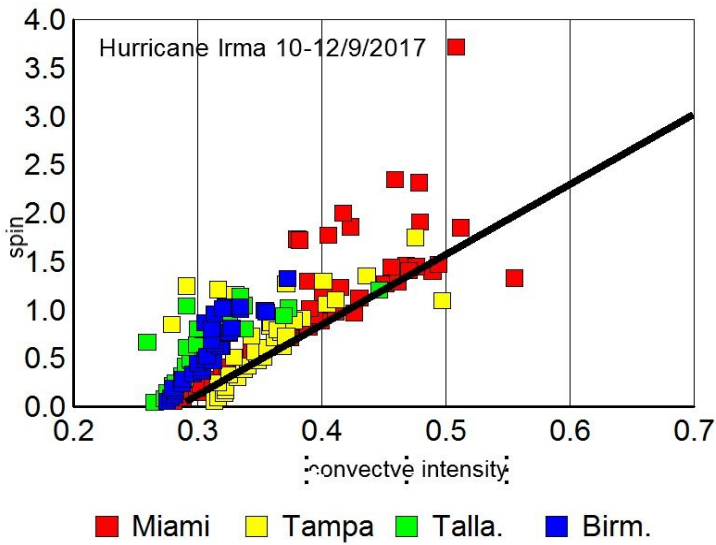

**Figure A4_2:** *Hurricane cluster spin for evolving convective intensity from 4 radar sites and that of possible imbedded tornadoes*.

If also true for a larger sample of tornadoes, it only remains a housekeeping task to identify the location of each tornado, measure its convectivity and spin, and its potential to do harm; and, for the significant tornadoes, track them, estimate their forward velocity and direction over several scans, and issue a public warning.

The Irma cluster data presented in Fig. A4_2 represents another dimension of the problem: that of variation with time of hurricane and tornado state. The data from the Miami radar on Sept. 10 represents the maximum extent and intensity of the cluster field. The data for Tampa Bay and Tallahassee early the next day shows less intensity, and that from the Birmingham radar captures Irma's conversion to an extra-tropical storm the next day.

Clearly the hurricane state estimated from the ($\mu$, $\Psi$) couplet reflects its evolution in two ways. The range of convectivity and spin is less the weaker the hurricane state. As well, there is a shift at in the variation of cluster spin variability: the more the convectivity the more its variation in spin. This suggests that as large clusters shrink or expand, their change is also controlled by another parameter, possibly the spatial variation of the rain field, an/or the removal of latent heating affects cluster size more than rotational speed. Also possible is that some clusters are tornadoes embedded in the hurricane field.

## *Acknowledgements:*

The author would like to thank Professors Emeritus Roger Pielke and William Cotton, Colorado State University, and Douw Steyn, University of British Columbia, for their support and encouragement, as well as the NOAA staff who helped provide him with the radar data.

Clarkson Univ., Potsdam, N.Y., 1998.

Kerman, B.R., P. Wadhams, N.R. Davis, and J. Comiso: Informational equivalence between synthetic aperture radar imagery and the thickness of Arctic pack ice. *J. Geophys. Res.*, 104(C12), 29721-9731, 1999.

Kleidon, A.: Non-equilibrium thermodynamics, maximum entropy production and Earth-system evolution. *Phil. Trans. R. Soc. A*, 368, 181–196, 2010.
https://doi.org/10.1098/rsta.2009.0188

Klotzbach, Michael M. Bell, Steven G. Bowen, Ethan J. Gibney, Kenneth R. Knapp, and Carl.J.Schreck III: Surface Pressure a More Skilful Predictor of Normalized Hurricane Damage than Maximum Sustained Wind. *BAMS.*, 103, 7, 2020.
https://doi.org/10.1175/bams-d-19-0062.1

Mali, P., S. K. Sarkar and J. Das: Raindrop size distribution measurements at a tropical station. *Indian Journal of Radio & Space Physics*. 32, 296-300, 2003.

Pauluis, O., and I. Held: Entropy Budget of an Atmosphere in Radiative–Convective Equilibrium. Part II: Latent Heat Transport and Moist Processes. *J. Atmos.. Sci.*, 59, pp 140-149, 2002.
https://doi.org/10.1175/1520-0469(2002)059%3c0125:eboaai%3e2.0.co;2

Press, W.H., B.P. Flannery, S.A. Teukolsky, and W.T. Vetterling: Numerical Recipes in Fortran. *Cambridge University Press*. 722P, 1989.

Richardson, L. F.: Atmospheric diffusion shown on a distance-neighbour graph. *Proc. Roy. Soc. Lond. A .,* 100, 709-737, 1926. https://doi.org/10.1098/rspa.1926.0043
41